# Self-assembly of flexible patchy nanoparticles in solution[*]

MENG Guoqing, CHEN Liyuan, GUO Sihang, PAN Junxing[*], WANG Yingying[*], ZHANG Jinjun[*]

School of Physics and Information Engineering, Shanxi Normal University, Taiyuan 030032, China

**Abstract**

The self-assembly of polymer grafted nanoparticles is more and more used in the field of functional materials. However, there is still a lack of analysis on the dynamic transformation paths of different self-assembly morphologies, which makes it impossible to achieve further precise regulation and targeted design in experiments and industrial production. In this work the effects of patchy property, grafted chain length, ratio and grafting density on the self-assembly behavior and structure of polymer grafted flexible patchy nanoparticles are investigated by dissipative particle dynamics simulation method through the construction of coarse-grained model of polymer grafted ternary nanoparticles. The influence and regulation mechanisms of these factors on the self-assembly structure transformation of flexible patchy nanoparticles are systematically studied, and a variety of structures such as dendritic structure, columnar structure, and bilayer membrane are obtained. The self-assembly structure of flexible patchy nanoparticles obtained in this work (such as bilayer membrane structure) provides a potential application basis for designing drug carriers. By precisely regulating the specific structural characteristics of the system, it is possible to achieve efficient loading of drugs and targeted delivery functions, thus significantly improving

---

the bioavailability and effect of drugs.

# 1. Introduction

Polymer-grafted nanoparticles refer to the modified nanoparticle structure formed by the modification of polymer chains onto the surface of inorganic nanoparticles through covalent bonds. If the polymer chain grafted to the surface of the nanoparticle is hydrophilic and hydrophobic, the resulting polymer-grafted nanoparticle in this case is amphiphilic. These nanoparticles and block copolymers have similar chemical properties, and they can form highly ordered structures through self-assembly in solution[1–7]. Polymer-grafted nanoparticles combine the unique physical and chemical properties of nanoparticles (such as optical properties[8] , magnetic properties[9] ,and catalytic properties, etc.) With the adjustability of polymers (such as flexibility, biocompatibility, etc.), so that they can have structures and functions that other materials can not achieve. Therefore, this nanocomposite has a wide range of potential applications in materials science, biomedical[10,11], energy[12] and other fields. Nanoparticles have poor dispersion in solvents[13], and it is difficult to form ordered aggregates by self-assembly. When the surface of these nanoparticles is grafted with polymer chains, the whole system can show controllable aggregation behavior under the joint action of various forces (interactions between nanoparticles, polymer chain-chain interactions, polymer chain-solvent interactions)[14–17]. Ordered structures with good properties (such as dendritic structure, columnar structure, bilayer membrane structure, etc.) Can be obtained by effectively and precisely controlling the self-assembly process of polymer-grafted nanoparticles.

With the rapid development of computer hardware and software, computer simulation technology has been widely used in the research of soft matter self-assembly because of its efficient and fast characteristics[18–24]. Computer simulation can not only provide design schemes for screening new materials, shorten the development cycle of new materials and reduce the development cost, but also further study the influence of individual factors on the self-assembly path transition of polymer-grafted nanoparticles, so as to discover the regulation mechanism of their self-assembly morphology. In recent years, there have been a large number of simulation studies on the self-assembly of polymer-grafted nanoparticles[25–36]. Song et al.[25] studied the self-assembly behavior of polymer-grafted nanoparticles in solution by dissipative particle dynamics simulation, and discussed in detail the effects of solvophobic segment length, solvophobic segment length ratio, solvent selectivity and other factors on the self-assembly behavior of the system, and obtained rich structures such as multi-chamber vesicles, onion-like vesicles, conchoidal micelles, double-layer micelles, etc. Ma et al.[28] Studied the self-assembly and aggregation behavior of amphiphilic block copolymer grafted nanoparticles in hybrid assemblies. By changing the number of arms of the grafted amphiphilic block copolymer, the length of the hydrophobic block, and the interaction parameters between the nanoparticles and the hydrophobic block, different morphologies of hybrid aggregates were obtained, including branched rod-like micelles, ring-like micelles, discoid micelles, and vesicles. Ma et al.[29] Also studied the self-assembly of shape amphiphiles composed of a hydrophilic head and one or more hydrophobic tails. By changing the interaction parameters between the hydrophilic head and the solvent, the length of the hydrophobic tail, the size of the hydrophilic head, and the number of hydrophobic tails, vesicles, discoidal micelles, and segmented rodlike micelles were obtained. Estridge and Jayaraman[30] studied the self-assembly behavior of AB diblock copolymer grafted particles in an implicit solvent by molecular dynamics simulation, and explored the effects of solvent selectivity, grafting length and particle size on the formation and assembly of plaque particles. Hou et al.[31] designed a new type of polymer chain grafted nanoparticles by coarse-grained molecular dynamics simulation, and discussed the effects of polymer chain length, density and stiffness on its self-assembly structure. So far, the research on the self-assembly of flexible polymer chain-grafted nanoparticles in solution has mainly focused on the influence of the properties of the grafted chain on the self-assembly structure of the system, while the influence of the central particle on the structure of the system has been less discussed[37–39]. Li et al.[37] studied the cooperative assembly of amphiphilic oligomers and Janus particles with different hydrophobic patch coverage. By changing the patch coverage, spherical, rod-like, disk-like and other self-assembly structures were obtained. Xu et al.[38] proposed a new non-spherical particle model. By using Brownian dynamics simulation method, various self-assembly structures such as vesicles, spherical micelles and so on were obtained by changing the aspect ratio and component ratio of ellipsoidal particles. These results indicate that the structure of the central particle plays an important role in the formation and regulation of the self-assembly structure of the system, and it is of great significance to further study the influence of the structure and properties of the central particle on the self-assembly structure.

In this paper, a coarse-grained model of flexible patchy nanoparticles was constructed, and the self-assembly behavior of the system was systematically studied by dissipative particle dynamics (DPD) simulation, focusing on the control mechanism of the hydrophilic/hydrophobic segment length ratio, system symmetry and grafting density on the self-assembly structure and dynamic behavior.

## 2. Model and method

The model used in this paper is shown in Fig. 1. The system is constituted by six different types of particles: hydrophilic segments made of white particles C with a length of $L_C$ and a grafting number of $N_C$, hydrophobic segments made of gray particles D with a length of $L_D$ and a grafting number of $N_D$, solvent particles S (not shown in Fig. 1), The middle patchy nanoparticle group is composed of particles with a diameter of 1. The patchy nanoparticle is composed of three parts, namely, a blue patch grafted with a hydrophilic segment, a yellow patch grafted with a hydrophobic segment, and a red patch not grafted with a segment.

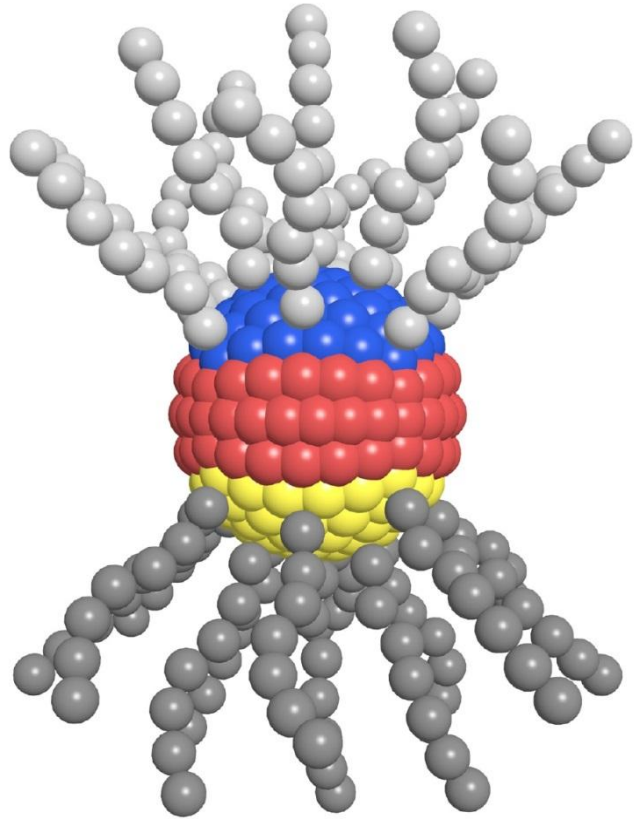

**Figure 1.** Polymer grafted nanoparticle model. The white and grey segments represent hydrophilic and hydrophobic segments, respectively. The blue and yellow patches correspond to the grafting regions of the hydrophil (C) and hydrophobic (D) chains, respectively, and the red patch corresponds to the non-grafting region.

In the simulation, the size of the simulation box is set to $L_X = L_Y = L_Z = 40$, 15 flexible patchy nanoparticles and solvent particles S are placed in the box, and the number density of all beads in the system is set to 3.0.

First proposed by Hoogerbrugge and Koelman[40] in 1992, DPD is a meshless particle-in-cell simulation algorithm for the simulation of complex fluid behavior at the mesoscopic scale. Compared with molecular dynamics (MD), it can simulate the hydrodynamic behavior of larger and more complex systems in the microsecond range[41–46]. In the DPD simulation, the coarse-grained DPD particles interact with each other through pairwise interaction forces, which include the conservative force $F^C$, the random force $F^R$ and the dissipative force $F^D$. All DPD particles obey Newton's equation of motion:

$$\frac{dr_i}{dt} = v_i, \frac{dv_i}{dt} = \frac{f_i}{m_i}, \quad (1)$$

Where $r_i$, $v_i$, $m_i$ represent the position, velocity, and mass of the *i* particle, respectively. $f_i$ represents the resultant force acting on the *i* particle, which is the sum of all pairwise interaction forces:

$$f_i = \sum_{j \neq i} (f_{ij}^C + f_{ij}^D + f_{ij}^R). \quad (2)$$

The three pairwise forces are

$$f_{ij}^C = a_{ij}\omega(r_{ij})\hat{r}_{ij}, \quad (3)$$

$$f_{ij}^D = -\gamma\omega^2(r_{ij})(\hat{r}_{ij} \cdot v_{ij})\hat{r}_{ij}, \quad (4)$$

$$f_{ij}^R = \sigma\omega(r_{ij})\theta_{ij}\hat{r}_{ij}. \quad (5)$$

Here $a_{ij}$ refers to the repulsive interaction parameter between particle *i* and particle *j*; $r_{ij} = r_i - r_j$, $r_{ij} = |r_{ij}|$, $\hat{r}_{ij} = r_{ij}/r_{ij}$; $v_{ij} = v_i - v_j$; $\gamma$ is the friction coefficient that controls the magnitude of the dissipative force; $\sigma$ is the noise amplitude controlling the strength of the random force; $\theta_{ij}$ is a normally distributed random variable. $\gamma$ and $\sigma$ are related by

$$\sigma^2 = 2\gamma k_B T. \quad (6)$$

The $\omega(r_{ij})$ weight function is derived from the work of Groot and Warren[47]. It is expressed as

$$\omega(r_{ij}) = \begin{cases} 1 - r_{ij}/r_c, & r_{ij} \leqslant r_c, \\ 0, & r_{ij} > r_c, \end{cases} \quad (7)$$

Where $r_c$ denotes the truncation radius. We performed a series of DPD simulations in the NVT ensemble, all with the help of the GPU-accelerated molecular simulation software PYGAMD[48–52]. The simulations were run with periodic boundary conditions, and each simulation was run for at least 5 × 10⁶ steps, the last 2 × 10⁶ steps were used for statistics, and the time step was set to $\Delta t = 0.04\tau$.

The interparticle interaction parameter can be expressed by the relationship between $\alpha_{ii}$ and the Flory-Huggins interaction parameter $\chi_{ij}$: $\alpha_{ij} = \alpha_{ii} + 3.497\chi_{ij}$. In the parameter setting, for the same DPD particles (i = A, B, C, D, E, S), we set the interaction parameter as $\alpha_{ii} = 25$, which means that the compatibility between the same particles is strong. The other interaction parameters are expressed in terms of symmetric matrices as follows:

$$\alpha_{ij} = \begin{bmatrix} & A & B & C & D & E & S \\ A & 25 & 27 & 50 & 27 & 50 & \alpha_{AS} \\ B & 27 & 25 & 35 & 25 & 35 & 29 \\ C & 50 & 35 & 25 & 35 & 25 & 35 \\ D & 27 & 25 & 35 & 25 & 35 & 29 \\ E & 50 & 35 & 25 & 35 & 25 & 35 \\ S & \alpha_{AS} & 29 & 35 & 29 & 35 & 25 \end{bmatrix}. \quad (8)$$

The larger the $\alpha_{ij}$ is, the stronger the repulsive force between the two particles is.

The $\alpha_{AS}$ in the matrix is the adjustment variable, which represents the force between the middle patch particle A and the solvent particle S. The adjustment range of the $\alpha_{AS}$ is 25-60, which corresponds to the change of the hydrophilic to hydrophobic properties of the A patch.

In this paper, the effects of $(\alpha_{AS})$, $(\Phi)$, $(L)$ and $(L_D)$ on the self-assembly behavior of flexible patchy nanoparticles in solution were investigated by selecting the patch properties $\alpha_{AS}$, grafting density $\Phi$, hydrophilic and hydrophobic segment length $(L)$ and hydrophobic segment length $L_D$ as the main parameters.

## 3. Results and Discussion

Four typical self-assembly structures were successfully obtained by controlling the relevant parameters: discrete structure (Fig. 2(a)), dendritic structure (Fig. 2(b)), columnar structure (Fig. 2(c)) and bilayer membrane structure (Fig. 2(d)). In the discrete structure, the flexible patchy nanoparticles have the highest dispersion, the interfacial contact area with the solvent is maximized, and the surface is coated by polymer chains. The dendritic structure is a highly bifurcated network formed by multiple flexible patchy nanoparticles connected end-to-end by the same graft chain, in which the contact area between the particle and the solvent is large, and the graft polymer chain mainly plays a "bridging" role. The columnar structure is a multi-layer stack, each layer is composed of three to five flexible patchy nanoparticles arranged in order, and the contact area between the nanoparticles and the solvent is small. Compared with the columnar structure, the bilayer membrane structure contains only two layers of nanoparticles, which has a symmetrical configuration. The overall performance is that the hydrophilic chain is outward and the hydrophobic chain is inward, and the contact area between the system and the solvent is minimized.

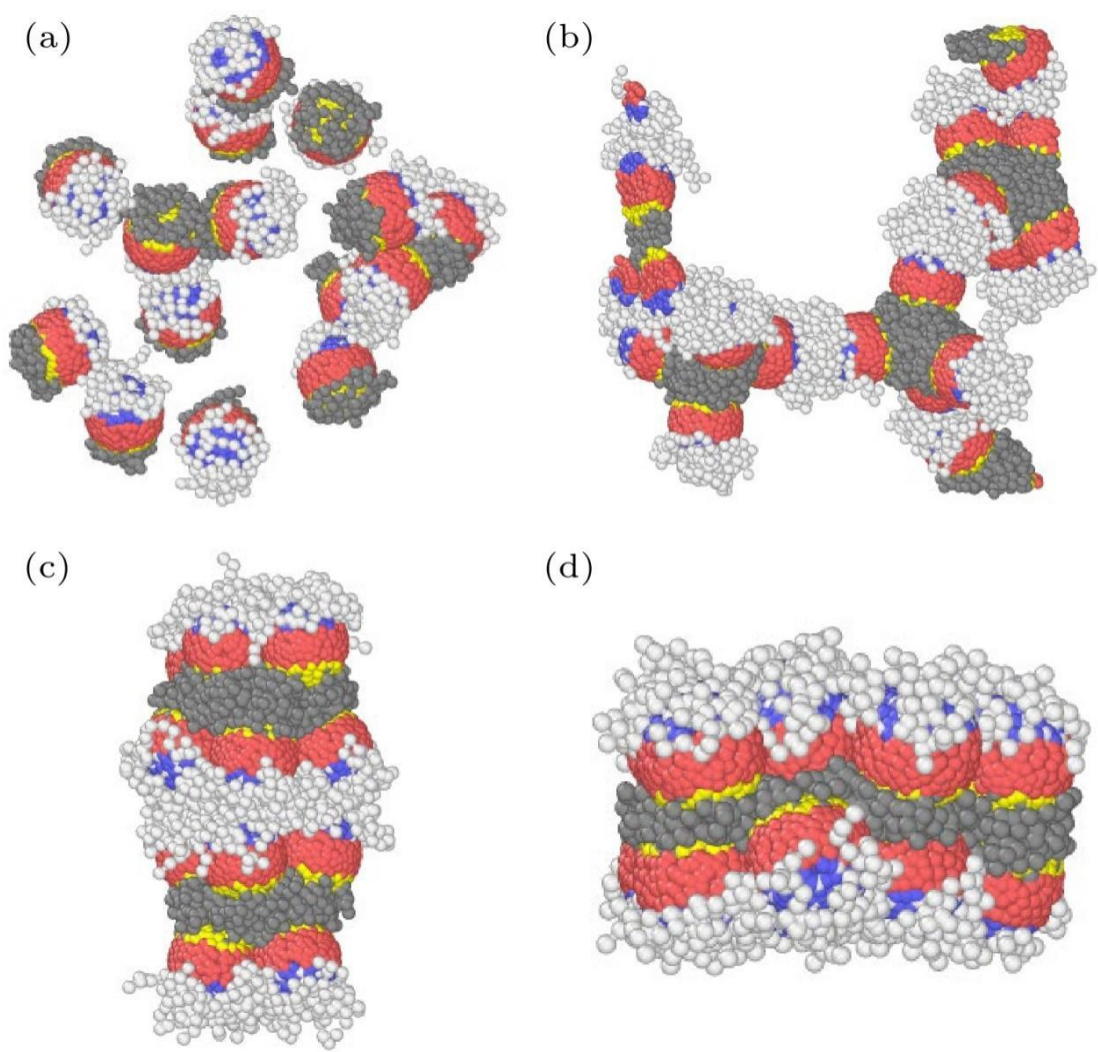

**Figure 2.** Typical structures during self-assembly processes of flexible patchy nanoparticles: (a) Discrete structure; (b) dendritic structure; (c) columnar structure; (d) bilayer membrane structure.

Firstly, under the condition of fixed graft chain number $N_C = N_D = 20$, the influence of the interaction force parameter $\alpha_{AS}$ between nanospheres and solvents and the graft chain length $L$ on the formation of the system structure was systematically discussed, where $L = L_C = L_D$. The specific results are shown in the Fig. 3. It can be seen that when the nanospheres are hydrophilic ($\alpha_{AS} = 25$), the nanospheres show a tendency to disperse in the system, and at the same time, due to the attraction of the grafted chains, the further dispersion of the nanospheres is limited, and finally the system shows a branched structure as a whole; As the repulsion between the nanospheres and the solvent increases, the nanospheres tend to aggregate closely to reduce the contact area with the solvent, thereby reducing the free energy of the system, and the structure of the system gradually changes from branched to columnar and bilayer membrane structures. On the other hand, the increase of the length of the grafted chain leads to the gradual increase of the volume occupied by the polymer chain on the nanosphere. At this time, in addition to the attraction between the polymer chains, the steric hindrance effect of the grafted chain becomes significant, which hinders the close packing between the particles, and finally leads to the gradual decrease of the formation region of the most closely arranged bilayer membrane structure in the phase diagram, while the second closely arranged columnar structure region increases. In general, the strong hydrophobicity and short chain length of nanoparticles contribute to the formation of bilayer membrane structure, while the long chain length contributes to the formation of columnar structure at a certain grafting density.

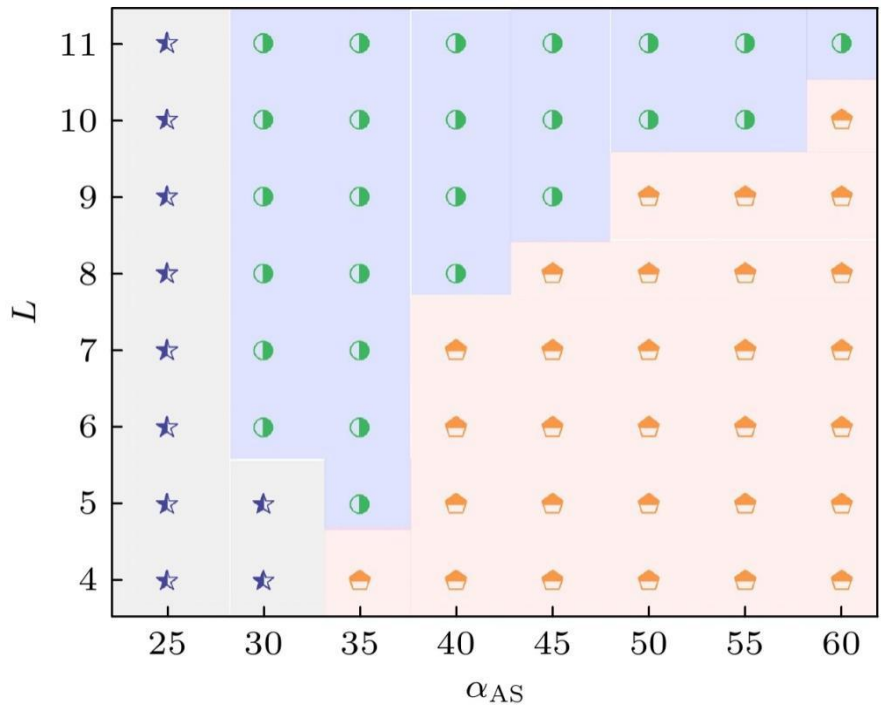

**Figure 3.** Phase diagram of flexible patchy copolymer self-assembly structures varying with the parameter of the driving force and the segment length $L$. ★ represents dendritic structure; ◐ represents a columnar structure; ⬠ represents a double-layer membrane structure.

Then, fixing the length of hydrophilic and hydrophobic chain segment $L_C = L_D = 7$, the effect of grafting density $\Phi$ on the structure formation of the system under different $\alpha_{AS}$ conditions was discussed, and the specific results are shown in the Fig. 4. Here, the grafting density $\Phi$ is calculated as $\Phi = N_g/S$, where $N_g$ is the total number of polymer chains grafted on the surface of the nanoparticle ($N_g = N_C + N_D$ and $N_C = N_D$), and $S$ is the area of the nanosphere grafting region. From the Fig. 4, it can be seen that under the condition of low grafting density ($\Phi < 0.65$), the system appears a discrete structure as shown in Fig. 2(a). This is because under the condition of low grafting density, the polymer chain is not enough to cover the grafting region of nanoparticles, and the "bridging" effect of the polymer is weak, which can not form a large aggregation structure. With the increase of the grafting density ($\Phi > 0.65$), the "bridging" effect of the grafted chains is enhanced, and the aggregation effect of the nanospheres begins to appear, resulting in a rich self-assembly structure of the system, such as dendritic, columnar, and bilayer membrane structures. At the same time, it can be found that increasing the grafting density is beneficial to the formation of bilayer membrane structure in solvophobic system. It is worth noting that from Fig. 4(b) and Fig. 4(c), it can be seen that the greater the grafting density, the greater the membrane thickness of bilayer membrane structure. This is due to the fact that the number of grafted chains increases with the increase of the grafting density in the system. The solvophobic segments aggregate under the action of the solvent in order to reduce the interfacial free energy. However, due to the limited volume occupied by each chain, the steric hindrance forces the interlayer spacing of the bilayer membrane to increase, and the macroscopic performance is that the membrane thickness increases.

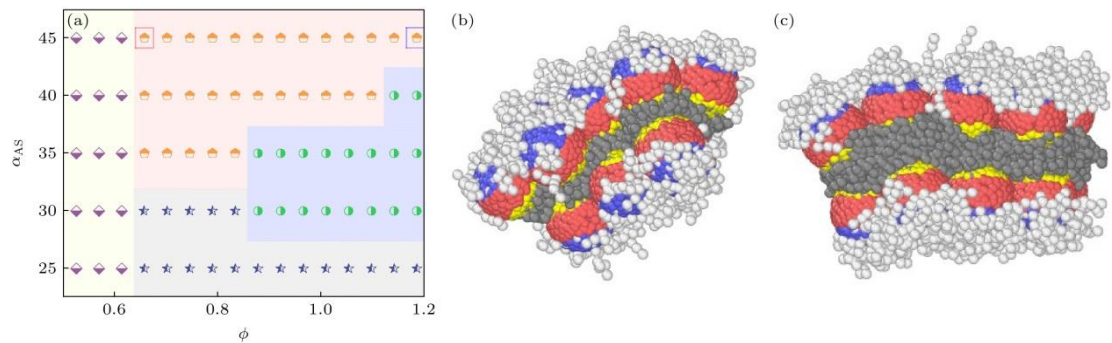

**Figure 4.** (a) Phase diagram of self-assembled structure of grafted nanoparticles with different grafting density polymers. ◆ represents discrete structure; ★ represents tree-like structure; ◐ represents columnar structure; ⬠ represents double-membrane

structure. (b) Schematic diagram of the structure of the double-layer membrane when $\Phi = 0.65$, $\alpha_{AS} = 45$. (c) Schematic diagram of the structure of the double- membrane when $\Phi = 1.18$, $\alpha_{AS} = 45$.

Then, when the number of grafted chains is $N_C = N_D = 20$ and the length of grafted hydrophilic segment is $L_C = 6$, the effect of the length of grafted hydrophobic segment $L_D$ on the formation of the system structure under different $\alpha_{AS}$ conditions is discussed, and the specific results are shown in the Fig. 5. It can be seen from the Fig. 5 that when the nanospheres are hydrophilic ($\alpha_{AS} = 25$), the solvent effect is more significant, and the system structure is dendritic no matter how the hydrophobic chain length is adjusted. When the repulsive force between the nanospheres and the solvent is strong ($\alpha_{AS} \geqslant 40$), the system shows a bilayer membrane structure, and the length of the hydrophobic segment has little effect on the formation of the system structure. However, when $25 < \alpha_{AS} < 40$, the bilayer membrane area increases with the increase of $L_D$, which is just opposite to the result in Fig. 3. Therefore, we counted the variation of the average membrane thickness $\overline{D}$ of the bilayer membrane structure formed by the system with the solvophobic chain length $L_D$ when $\alpha_{AS} = 40$, as shown in Fig. 6. It can be seen from Fig. 6 that the average membrane thickness of the bilayer membrane structure is changed by the growth of the hydrophobic segment, and the average membrane thickness $\overline{D}$ increases approximately linearly with the increase of the hydrophobic segment. It can also be seen from the inset of Fig. 6 that the increase of hydrophobic segment length not only increases the film thickness, but also combs and adjusts the regularity of the bilayer membrane. The corresponding aggregate morphology of the system is also given in the figure.

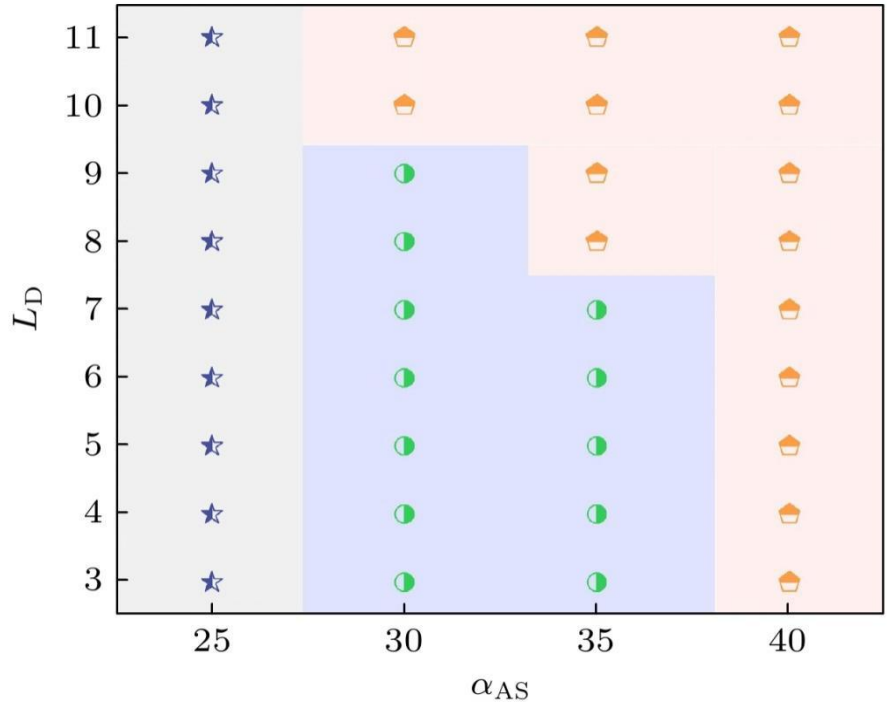

**Figure 5.** Phase diagram of the influence of solvent-swollen segment length on polymer morphology. ★ represents tree-like structure; ◐ represents columnar structure; ⬠ represents double-membrane structure.

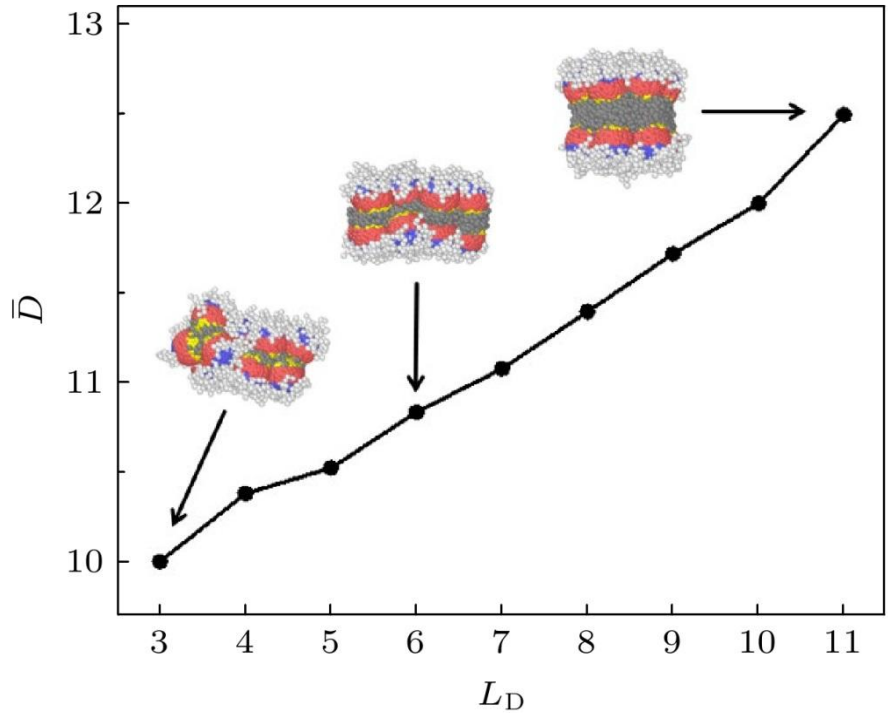

**Figure 6.** Variation of the average membrane thickness of the bilayer structure with the length of the hydrophobic solvent chain.

It is worth noting that the number of branches appearing in the Fig. 3 — Fig. 5 is also related to the size of the system. The results show that the number of branches of the dendritic structure increases with the increase of the size of the system (the side length of the box and the number of particles). However, due to the sensitivity of the dendritic structure to the initial conditions, the quantitative relationship between the system size and the number of branches remains to be further studied.

In order to quantitatively distinguish different typical self-assembly structures, the radial distribution function (RDF) of the central particle was calculated. The definition of RDF is $g(r) = \frac{d_n(r)}{4\pi r^2 \Delta r \rho}$, where $d_n(r)$ is the average of the total number of particles around the reference particle from $r \to r + \Delta r$ within the shell volume, $\rho = N/V$ is the number of particles per unit volume, $V$ is the volume of the system, and $N$ is the total number of particles in the system. We selected $L_C = L_D = 8$, the number of grafted chains $N_C = N_D = 20$, $\alpha_{AS} = 25, 40, 60$ to demonstrate the dendritic structure, columnar structure and bilayer membrane structure, respectively. The radial distribution function curves of three typical structures are given by Fig. 7. It can be seen from Fig. 7 that when $\alpha_{AS} = 25$, the first peak corresponds to $g(r)_{peak} \approx 11.4$ and $r_{peak} \approx 6.3$;

when $\alpha_{AS} = 40$ and $\alpha_{AS} = 60$, the first peak corresponds to $g(r)_{peak}$ of about 36.9 and 109.4, respectively, and the corresponding $r_{peak}$ of about 6.1 and 5.9. As can be concluded, with the increase in $\alpha_{AS}$, the peak intensity g corresponding to the first peak increases progressively, while the associated $r_{peak}$ shifts continuously toward the origin. A smaller $r_{peak}$ value indicates a more compact packing of particles, and a larger $g(r)_{peak}$ value corresponds to a greater inward contraction of the assembled structure. Compared with the first peak, the difference between the columnar structure and the layered structure is larger, indicating that the overall structure of the system tends to shrink; However, the two peak $g(r)_{peak}$ of the dendritic structure are approximately equal, indicating that the architecture tends to be decentralized. Secondly, from the number of peaks, the number of peaks in the bilayer membrane structure is the least, which indicates that all particles are confined in the bilayer membrane; The columnar structure takes the second place, indicating that it corresponds to the multi-layer structure of the system; The dendritic structure shows multiple low peaks, corresponding to its loose packing characteristics. From the peak width, the peak width reflects the degree of dispersion of the distance distribution between particles. It is observed from Fig. 7 that the peak width of the radial distribution function corresponding to the dendritic structure is the largest, indicating that the distance between particles in the structure fluctuates greatly and the distribution is uneven. In contrast, the peak width of the columnar structure and the bilayer membrane structure is smaller, and the smaller peak width indicates that the distance between particles is more uniform and the distribution is more concentrated.

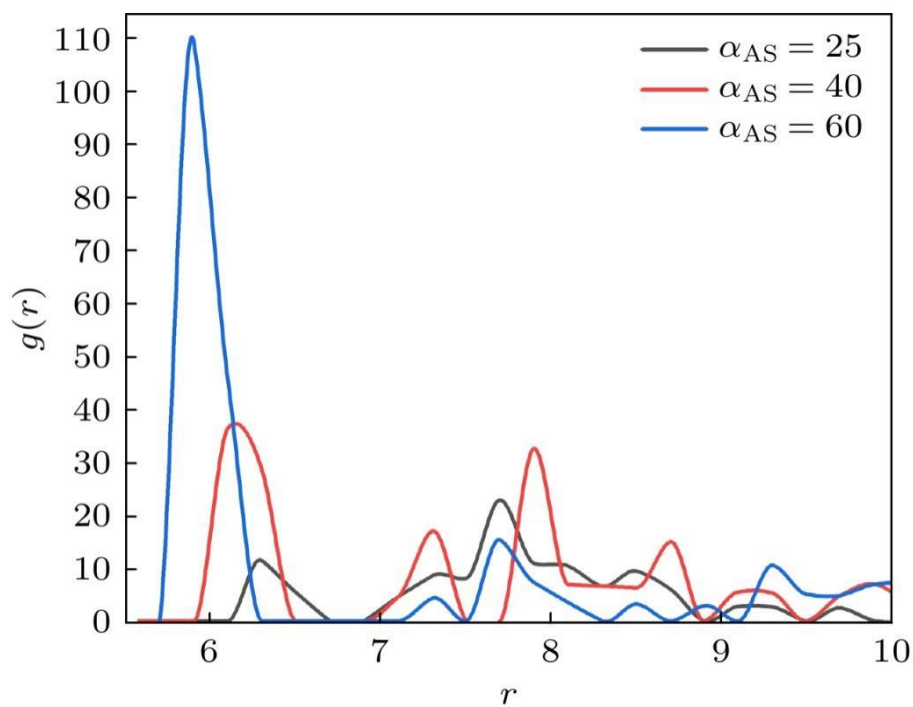

**Figure 7.** Rdial distribution function $g(r)$ of the central particle in the system. Black line represents the dendritic structure; red line represents the columnar structure; blue line represents the bilayer membrane structure.

In order to further explore the dynamic mechanism of architecture formation, we tracked and analyzed the evolution paths of three typical architectures, and Fig. 8 gave the evolution topography of the architecture with time. For the dendritic structure, the system undergoes a transition from a dispersed short rod-like structure to a ribbon-like structure and then to a dendritic structure during the formation process, which is mainly driven by the attraction between the grafted chains; For the columnar structure, the system undergoes a process from a short rod-like structure to a polyhedral structure, then to a cluster structure, and finally to a columnar structure under the concerted regulation of the interaction between the graft chains and the solvent; For the bilayer membrane structure, it first changes from a short rod-like structure to a cluster structure, and then goes through a long period of internal structure reorganization and relaxation, and finally becomes a bilayer membrane structure. This lengthy structural adjustment is the result of the synergistic effect of the attractive interaction between the grafted chains, the nanosphere-solvent interaction, and the "steric hindrance" effect.

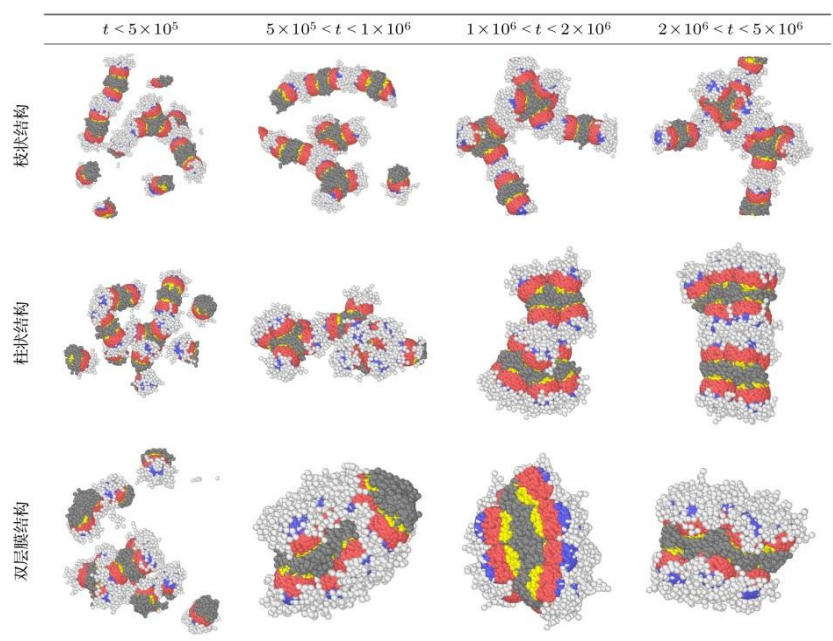

**Figure 8.** Different conformations of the self-assembly process of branched, columnar, and double-layer membrane structures.

By comparing the self-assembly process of dendritic, columnar and bilayer membrane structures, it is found that the path switching in the initial stage plays a key role in the final conformation. The length of the grafted segment and the solvent environment synergistically regulate the coiled state of the hydrophobic segment and the extended degree of the hydrophilic segment. This regulatory mechanism directly affects the interfacial contact area between the grafted chain segment and the solvent and the local interfacial curvature, which in turn determines the final self-assembly conformation of the flexible patchy nanoparticles.

In this study, the effect of the middle patch on the self-assembly structure of the system is considered, which is different from the previous researchers who mostly focused on the effect of the nature of the graft chain on the self-assembly structure of the system. We used the patchy ternary nanoparticles to construct a rigid-flexible alternative particle model, and through the self-assembly of these particles in solution, we obtained dendritic, columnar, bilayer membrane and other structures, which also showed the rigid-flexible alternative structural characteristics. These structures effectively combine the advantages of rigid and flexible particles, and are expected to achieve special properties that non-alternating nanoparticles do not have, which will provide new ideas for the design of nano-functional materials.

## 4. Conclusion

In this paper, a coarse-grained model of flexible patchy nanoparticles is constructed, and dissipative particle dynamics simulations are used to systematically study the effects of patch properties, length, proportion and grafting density of grafting chains on their self-assembly behavior and structure. The results show that the self-assembly of flexible patchy nanoparticles can form a variety of morphologies, such as dendritic, columnar and bilayer membrane structures. When the compatibility between the solvent and the plaque changes, the self-assembled structure will change from a dendritic structure to a columnar structure or a bilayer membrane structure. In addition, the increase of chain segment length can also cause a similar transformation of the assembly structure. Further studies have shown that the grafting density of nanoparticles also has an important effect on the self-assembly structure. With the increase of grafting density, the self-assembly structure of flexible patchy nanoparticles exhibits discrete structure, dendritic structure and bilayer membrane structure in turn when the solvent interaction parameter remains unchanged. In addition, the length of the hydrophobic segment also has a significant effect on the self-assembly structure of the system. Under the condition of suitable interaction parameters, with the increase of the length of the hydrophobic segment, the region of the system forming a bilayer membrane structure expands accordingly.

It is worth noting that in the simulation process, the solvent is simplified as a single parameter, while the chemical properties and physical States of the solvent in the actual system are often more complex. This simplification may not fully reflect the actual role of solvent in the self-assembly process, especially the self-assembly behavior under complex conditions still needs further exploration. Future research is expected to develop more refined models, such as the introduction of solvent polarity, temperature dependence and specific interactions between solvent and polymer, so as to more accurately capture the conformational details of polymer chains and the dynamic behavior of solvent molecules, which will provide strong support for more accurate simulation of self-assembly processes in real systems.

## References

[1]  Hoheisel T N, Hur K Wiesner U B 2015 *Prog. Polym. Sci.* **40** 3
[2]  MaiY Y, Eisenberg A 2012 *Acc. Chem. Res.* **45** 1657
[3]  Yu L X Z, Shi R, Qian H J, Lu Z Y 2019 *Phys. Chem. Chem. Phys.* **21** 1417
[4]  Maiorov B, Baily S A, Zhou H, Ugurlu O, Kennison J A, Dowden P C, Holesinger T G, Foltyn S R, Civale L 2009 *Nat. Mater.* **8** 398
[5]  Yang J Y, Hu Y, Wang R, Xie D Q 2017 *Soft Matter* **13** 7840

[6]  Li Q, Wang L Y, Lin J P, Zhang L S 2019 *Phys. Chem. Chem. Phys.* **21** 2651

[7]  Li Q, Wang L Q, Lin J P, Xu Z W 2020 *J. Phys. Chem. B* **124** 4319

[8]  Varanakkottu S N, Anyfantakis M, Morel M, Rudiuk S, Baigl D 2015 *ACS Nano* **16** 644

[9]  Zhang X, Yang H, Xiong H M, Li F Y, Xia Y Y 2006 *J. Power Sources* **160** 1451

[10] Han X, Xue B, Cao Y, Wang W 2024 *Acta Phys. Sin.* **73** 178103

[11] Song J B, Zhou J J, Duan H W 2012 *J. Am. Chem. Soc.* **134** 13458

[12] Wang Q, Li Z Y, Deng H Y, Chen Y G, Yan Y G 2023 *Chem. Commun.* **59** 6726

[13] Bansal A, Yang H C, Li C Z, Cho K W, Benicewicz B C, Schadler L S 2005 *Nat. Mater.* **4** 693

[14] Mackay M E, Tuteja A, Duxbury P M, Hawker C J, Van Horn B, Guan Z B, Chen G H, Krishnan R S 2006 *Science* **311** 1740

[15] Shao M X, Lin J P, Zhang L S 2025 *J. Funct. Polymers* **38** 318

[16] Krishnamoorti R 2007 *MRS Bull.* **32** 341

[17] Wang Y Y, Chen L Y, Lu J F, Pan J X, Zhang J J 2024 *Langmuir* **40** 16595

[18] Duan F L, Wang Y 2014 *Acta Phys. Sin.* **63** 136102

[19] Götz A W, Clark M A, Walker R C J 2014 *J. Comput. Chem.* **35** 95

[20] Pecina A, Lepšík M, Řezáč J, Brynda J, Mader P, Řezáčová P, Hobza P, Fanfrlík J 2013 *J. Phys. Chem. B* **117** 16096

[21] Li H, Zhao H T, Gao K M, Xue Z J, Chen Z B, Liu H 2024 *Polym. Int.* **74** 152

[22] Li C H, Fu X W, Zhong W H, Liu J 2019 *ACS Omega* **4** 10216

[23] Pal S, Seidel C 2006 *Macromol. Theory Simul.* **15** 668

[24] Zhong C L, Liu D H 2007 *Macromol. Theory Simul.* **16** 141

[25] Song W Y, Liu H, He J W, Zhu J Z, He S Y, Liu D H, Liu H, Wang Y 2022 *Polym. Int.* **71** 1330

[26] Liu H, Zhao H Y, Müller-Plathe F, Qian H J, Sun Z Y, Lu Z Y 2018 *Macromolecules* **51** 3758

[27] Xing J Y, Lu Z Y, Liu H, Xue Y H 2018 *Phys. Chem. Chem. Phys.* **20** 2066

[28] Ma S Y, Hu Y, Wang R 2016 *Macromolecules* **49** 3535

[29] Ma S Y, Hu Y, Wang R 2015 *Macromolecules* **48** 3112

[30] Estridge C E, Jayaraman A 2014 *J. Chem. Phys.* **140** 144905

[31] Hou G Y, Xia X Y, Liu J, Wang W C, Dong M J, Zhang L Q 2019 *J. Phys. Chem. B* **123** 2157

[32] Gupta S, Chokshi P 2020 *J. Phys. Chem. B* **124** 11738

[33] Moinuddin M, Tripathy M 2022 *Macromolecules* **55** 9312

[34] Sriramoju K K, Padmanabhan V 2016 *Macromol. Theory Simul.* **25** 582

[35] Li L, Han C, Xu D, Xing J Y, Xue Y H, Liu H 2018 *Phys. Chem. Chem. Phys.* **20** 18400

[36] Shi R, Qian H J, Lu Z Y 2017 *Phys. Chem. Chem. Phys.* **19** 16524

[37] Li J W, Wang J F, Yao Q, Yu K, Zhang J 2019 *Nanoscale* **11** 7221

[38] Xu J, Wang Y L, He X H 2015 *Soft Matter* **11** 7433

[39] Yan L T, Popp N, Ghosh S K, Böker A 2010 *ACS Nano* **4** 913

[40] Hoogerbrugge P J, Koelman J M V A 1992 *Europhys. Lett.* （EPL） **19** 155

[41] Park S, Lee J H, Cho M, Lee Y S, Chung H, Yang S 2024 *Polym. Test.* **137** 108531

[42] Zhang C, Wang Z G, Wang X H, Mou X K, Li S B 2024 *Polymer* **312** 127664

[43] Dong H, Zhou H, Li Y F, Li X B, Fan L L, Wen B Y, Zhao L 2024 *Macromol. Theory Simul.* **34** 2400078

[44] Zhu B W, He Z J, Jiang G S, Ning F L 2024 *Polymer* **290** 126602

[45] Ma Y B, Yuan X Q, Jiang R F, Liao J H, Yu R T, Chen Y P, Liao L S 2023 *Polymers* **15** 856

[46] Wang F, Feng L K, Li Y D, Guo H X 2023 *Chin. J. Polym. Sci.* **41** 1392

[47] Groot R D, Warren P B 1997 *J. Chem. Phys.* **107** 11

[48] Hu F F, Sun Y W, Zhu Y L, Huang Y N, Li Z W, Sun Z Y 2019 *Nanoscale* **11** 17350

[49] Li Z W, Zhu Y L, Lu Z Y, Sun Z Y 2016 *Soft Matter* **12** 741

[50] Zhu Y L, Pan D, Li Z W, Liu H, Qian H J, Zhao Y, Lu Z Y, Sun Z Y 2018 *Mol. Phys.* **116** 1065

[51] Zhu Y L, Liu H, Li Z W, Qian H J, Milano G, Lu Z Y 2013 *J. Comput. Chem.* **34** 2197

[52] Wang Y H, Zhou Q Z, Zhu Y L, Fu C L, Huang Y N, Li Z W, Sun Z Y 2019 *Chem. J. Chin. Univ.* **40** 1037